\documentclass{elsart}
\usepackage{graphics}
\def\normord#1{\mathopen{\hbox{\bf:}}#1\mathclose{\hbox{\bf:}}}
\def\ha{{1\over 2}}
\def\del{\partial}
\def\lan{\langle}
\def\ran{\rangle}
\def\bra#1{\lan#1|}
\def\ket#1{|#1\ran}

\def\ra{\rightarrow}
\def\frac#1,#2{{#1\over #2}}

\def\fr#1,#2{{#1\over #2}}

\begin{document}
\begin{frontmatter}
\hyphenation{Coul-omb ei-gen-val-ue ei-gen-func-tion Ha-mil-to-ni-an
  trans-ver-sal mo-men-tum re-nor-ma-li-zed mas-ses sym-me-tri-za-tion
  dis-cre-ti-za-tion dia-go-na-li-za-tion in-ter-val pro-ba-bi-li-ty
  ha-dro-nic he-li-ci-ty Yu-ka-wa con-si-de-ra-tions spec-tra
  spec-trum cor-res-pond-ing-ly}
\title{Chiral Symmetry {\it Breaking} in the\\ Light-Cone Representation\\ 
($M_\pi^2 \sim \mu_q \langle\Omega |\bar{\psi}\psi|\Omega \rangle +\dots$)}
\author{Gary McCartor }
\address{Dept. of Physics, SMU, Dallas, TX 75275,\\
         mccartor@mail.physics.smu.edu}
\date{19 June 2000}
\begin{abstract}
In this paper I shall discuss the way in which vacuum structure and condensates occur in the light-cone representation.  I shall particularly emphasize the mechanism by which the mass squared of a composite such as the pion comes to depend linearly on the bare mass of its Fermion constituents.  I shall give details in two dimensions then discuss the case of four dimensions more speculatively.
  \end{abstract}
\end{frontmatter}
\section{Introduction}
\label{introduction}
At the Heidelberg meeting in 1991 there was much talk of zero modes.  I was one of the ones talking about zero modes but there were many others.  The zero modes were discussed with regard to the puzzles: how can vacuum structure be represented? how can condensates form? how can one get a linear dependence of the physical-mass-squared of a composite like the pion or the Schwinger particle on the bare mass of the Fermion?  The zero modes also received discussion in other contexts.  Indeed there may have been some thought that the main problem facing the community was properly to include all necessary zero modes.

Now, in 2000, I have come to see the question not as one of zero modes, but as: how should one regulate the $P^+ = 0$ singularity?  Zero modes are often involved in the answer to that question, but the central question is properly regulating the singularity.  In two dimensions we can now can give answers to the above questions in full: we know how the vacuum gets dressed; we know how condensates form; we understand the difference between the continuum and the discretized cases and we know the operator which gives the linear dependence of the composite mass squared on the bare mass.  In four dimensions we know far less but we do understand some things.

In the next section I shall discuss the case of the Schwinger model. That will let me illustrate some of the above remarks and set a framework for the things I really want to say.  In particular, I will discuss vacuum structure and the chiral condensate.

I shall them move on the case of the massive Schwinger model.  There I shall concentrate on how the mass squared of the Schwinger particle depends linearly on the mass of the electron; that is, on the relation $m_S^2 = {e^2 \over \pi} - 4 \pi\mu \bra{\Omega}\bar{\Psi}\Psi\ket{\Omega} + \dots$

Finally, I shall move on to four dimensions, particularly the case of QCD.  At that point the discussion will become considerably more speculative.

\section{The Schwinger Model}
The Lagrangian for the massive Schwinger model is
$$
 {\mathcal {L}} =  i \bar{\psi} \gamma^{\mu}
\partial_{\mu} \psi 
- \fr{1},{4} F^{\mu \nu} F_{\mu \nu} 
-  A^{\mu} {J}_{\mu} -  \lambda A^+ - \mu\bar{\psi}\psi
$$
In this section we shall consider the Schwinger model ($\mu = 0$).  For that case the solution is~\cite{nm00}
$$
     \Psi_+ = Z_+ e^{\Lambda_+^{(-)}}\sigma_+ e^{\Lambda_+^{(+)}}
$$
$$
      \Lambda_+ = -i2\sqrt{\pi}({\eta}(x^+) + \tilde{\Sigma}(x^+,x^-))
$$
$$
       Z_+^2 = \fr{m^2e^\gamma},{8\pi\kappa}
$$
$$
     \Psi_- = Z_-e^{\Lambda_-^{(-)}}\sigma_- e^{\Lambda_-^{(+)}}
$$
$$
        Z_-^2 = \fr{\kappa e^\gamma},{2\pi}
$$
$$
      \Lambda_- = -i2\sqrt{\pi}\phi(x^+)
$$
$$
      \lambda =m{\partial }_{+}({\eta }-{\phi }) 
$$
$$
        A_+ = \fr{2},{m} \partial_+ ({\eta} + \tilde{\Sigma})
$$
The solution is written in terms of three free fields which are described as
$$
   \tilde{\Sigma}~ - ~ MASSIVE(m = \fr{e},{\sqrt{\pi}})~~PSEUDOSCALAR
$$
$$
 \phi~ - ~CHIRAL~SCALAR~;~~ \eta ~ - ~ CHIRAL~GHOST
$$
The massless fields are regulated using the Klaiber~\cite{kla} procedure.  For instance
$$
\phi^{(+)}(x^{+})=i(4\pi )^{-\ha}\int_{0}^{\infty
}dk_{+}k_{+}^{-1}d(k_{+})\left({\rm e}^{-ik_+ x^+}-\theta 
(\kappa -k_{+})\right)
$$
It is important to realize that the solution is in light-cone gauge but not light-cone quantized.  As written, it is representation independent.  It can be found using equal-time quantization, light-cone quantization or by quantizing on $x^- = 0$\cite{nm00}.  Unless care is taken, the massless fields, $\phi$ and $\eta$, are likely to be missed in standard light-cone quantization.  Therefore I wish to spend some time discussing the role these unphysical fields play in the solution.

The massless fields are needed to maintain covariance and to regulate the $P^+ = 0$ singularity.  Consider the Fermi product for the $\Psi_+$ field.  First, we note that the product canot be regulated by splitting in the $x^-$ direction.  That is really the major complication in finding the solution using light-cone quantization.  I shall return to that point in the discussion of the massive Schwinger model.  For the solution to work, the singularity in the Fermi two point function must be the same as for free fields
$$
    \langle \Psi_+^*(x+\epsilon)\Psi_+(x)\rangle\sim \fr{1},{2\pi\epsilon^-}
$$
where $\epsilon$ is space-like.  But if we include only the physical pseudoscalar we find
$$
   \langle :e^{i2\sqrt{\pi}\tilde{\Sigma}(x + \epsilon)}::e^{-i2\sqrt{\pi}\tilde{\Sigma}(x)}:\rangle \sim e^{-2\gamma}\fr{4},{m^2} \fr{1},{\epsilon^+\epsilon^-}
$$
Things are put right by the contribution from the ghost
$$
\langle e^{i2\sqrt{\pi}\eta^{(-)}(x+\epsilon)}\sigma_+^*e^{i2\sqrt{\pi}\eta^{(+)}(x+\epsilon)}e^{-i2\sqrt{\pi}\eta^{(-)}(x)}\sigma_+e^{-i2\sqrt{\pi}\eta^{(+)}(x)}\rangle\sim e^{\gamma}\kappa \epsilon^+
$$
With all this in place we can define a gauge invariant Fermi product as
$$
\normord{\psi^\dagger\psi}=\lim_{\e\ra0\atop \e^2<0}\Biggl\{
	e^{-ie\int_x^{x+\e} A_\nu^{(-)}dx^\nu}
	\psi^\dagger(x+\e)\psi(x)e^{-ie\int_x^{x+\e} A_\nu^{(+)}dx^\nu}
		-{\rm V.E.V.}\Biggr\}
$$
As far as I know, no gauge invariant, covariant regulator for the Schwinger model has been given which does not involve the use of a ghost field.

Of special interest to us will be the spurion operators; these operators are made up entirely of modes from the massless fields
$$
\sigma_+ ={\rm exp}\Big[ i\sqrt{\pi}\{Q_5 + Q\}(4m)^{-1} +\int_{0}^\kappa dk_1
k_1^{-1}\{\eta(k_1)-\eta^*(k_1)\}\Big]
$$
$$
 \sigma_- ={\rm exp}\Big[ i\sqrt{\pi}\{Q_5 - Q\}(4m)^{-1} + \int_0^{\kappa} dk_1
k_0^{-1}\{\phi(k_1) - \phi^*(k_1)\}\Big]  
$$
$$
  \sigma^*_\pm = \sigma_\pm^{-1}
$$
The charges are also composed entirely of modes from the massless fields
$$
    Q= \int_{-{\infty}}^{\infty}m{\partial}_+({\phi}-{\eta})dx^+        
$$
$$
    Q_5= \int_{-{\infty}}^{\infty}m{\partial}_+({\phi}+{\eta})dx^+
$$
Indeed, in the representation we are using, the spurions are the only operators which carry a charge or a pseudocharge
$$
[Q,\sigma^*_+\sigma_-] = [Q,\sigma^*_-\sigma_+] = 0
$$
$$
[Q_5,\sigma^*_+\sigma_-] = -2\sigma^*_+\sigma_-~;~ [Q_5,\sigma^*_-\sigma_+] = 2\sigma^*_-\sigma_+
$$
All other independent operators commute with $Q$ and $Q_5$.

We have seen that the massless, zero-mode fields are necessary to regulate the theory and preserve covariance and gauge invariance.  Now we want to see that once they are in place, they necessarily lead to the $\theta$-vacuum structure and to the chiral condensate.  The physical subspace for the Schwinger model is defined by the completion of
$$
                   \lambda^{(+)}\ket{S} = Q \ket{S} = 0   
$$
The completion of this space, $\bar{S}$, includes translationally invariant states which are not in the space of a free, massless Fermi field~\cite{str}.  These states all differ from the bare vacuum by sums of zero norm vectors.  They are all given by
$$
  |\Omega (M)\rangle =(-{\sigma }_{+}^{\ast }{\sigma }_{-})^{M}|0\rangle
$$

When we form the physical Hilbert space by factoring out the zero norm states in the physical space: ${\mathcal {P}} \equiv S/{\mathcal {K}}$, where ${\mathcal {K}}$ is the subspace of zero norm states, we must choose a state to represent the class which includes the bare vacuum; that will be the physical vacuum of the theory.  Gauge invariance requires that this state be an eigenstate of the generators of residual gauge invariance which are the chargeless combinations of spurions
$$
   (\sigma_+^*\sigma_-)~ A_+~(\sigma_+\sigma_-^*) =A_+ + \fr{\sin{\kappa x^+}},{\kappa}    
$$
Thus the physical vacuum must be of the form
$$
|\Omega (\theta )\rangle \equiv \sum_{M=-\infty }^{\infty }{\rm e}^{iM\theta
}|\Omega (M)\rangle 
$$
In that state we find the chiral condensate to be
$$
 \langle\Omega (\theta )|\bar{\Psi}\Psi|\Omega (\theta )\rangle = -{\frac{m},{2\pi}}
{\rm e}^\gamma \cos\theta
$$
\section{Adding a Mass}
We now consider the effect of a nonzero value of $\mu$.  We shall concentrate on the question: How can we have $M(\mu)^2 = M(0)^2 - 4 \pi \mu\langle\Omega (\theta )|\bar{\Psi}\Psi|\Omega (\theta )\rangle\dots$?  The Lagrangian will now include the term
$$
           \delta{\mathcal {L}} = - \mu\bar{\Psi}\Psi
$$
This induces an additional term in $P^-$
$$
        \delta P^- = \mu\int \bar{\Psi} \Psi = \mu\int (\Psi_-^*\Psi_+ + \Psi_+^*\Psi_-)
$$
To evaluate this expression we must solve the constraint equation
$$
{\del\Psi_-\over\del x^-} + i\ha\mu\Psi_+ =0 
$$
The solution is
$$
   \Psi_- = -\int i\ha\mu\Psi_+ dx^-
$$
but which particular antiderivative should we take.  What has usually been done in the past is to define the antiderivative as follows: if
$$
    \Psi_+(x^-) = \int dk_-
   (b(k_-) e^{-ik_-x^-} + 
   d^*(k_-) e^{ik_-x^-})        
$$
then we shall take ``$\int$'' to mean
$$
    \int \Psi_+(x^-) = \int dk_-
   (\fr{1},{-ik_-}b(k_-) e^{-ik_-x^-} + 
   \fr{1},{ik_-}d^*(k_-) e^{ik_-x^-} )       
$$
If this expression is used as the solution to the constraint equation then we find, at least naively, that $\delta P^- \sim \mu^2$ which leads to the puzzle as to how we can have $M(\mu)^2 = M(0)^2 - 4 \pi \mu\langle\Omega (\theta )|\bar{\Psi}\Psi|\Omega (\theta )\rangle\dots$.

But the general solution to the constraint equation is
$$
   \Psi_- = \Psi_-^0 -\int i\ha\mu\Psi_+ dx^-
$$
where $\Psi_-^0$ is any function of $x^+$.  From our solution to the Schwinger model we see that 
$\Psi_-^0$ is not zero since at $\mu = 0$ we have
$$
     \Psi_- = \Psi_-^0 = Z_-e^{\Lambda_-^{(-)}}\sigma_- e^{\Lambda_-^{(+)}}
$$
So we must have
$$
     \lim_{\mu\rightarrow 0} \Psi_-^0 = Z_-e^{\Lambda_-^{(-)}}\sigma_- e^{\Lambda_-^{(+)}}
$$
For general values of $\mu$ the solution is~\cite{mccartor99}
$$
         \Psi_-^0 = Z_-(\mu)e^{\Lambda_-^{(-)}(\mu)}\sigma_- e^{\Lambda_-^{(+)}(\mu)}
$$
where
$$
         \Lambda_-^{(-)}(\mu) = i(4\pi )^{-\ha} \int_{0}^{\infty
}{dk_{+} \over k_{+}}({\phi}(k_{+}) + f(\mu,k_+) {\lambda}(k_+))\left({\rm e}^{-ik_+ x^+}-\theta 
(\kappa -k_{+})\right)
$$
and the functions $f(0,m,k_+) = 0$ and $Z_-(\mu)$ are determined by the requirement that the full $\Psi_-$
$$
\Psi_- = \Psi_-^0 + \ha\mu^2 \int_{-\infty}^\infty \left(\Psi_+^*(0,x^-) \left[-\int i\ha \Psi_+ \right] + C.C.\right)dx^-
$$
satisfy the relation
$$
 \{\Psi_-(x),\Psi_-(x + \epsilon_{SPACE})\} = \delta(\epsilon_{SPACE})
$$
From our solution to the Schwinger model we know that 
$$
        f(0,m,k_+) = 0\quad{\rm and}\quad Z_-^2(0) = \fr{\chi e^\gamma},{2\pi}
$$
When we construct the operator $\bar{\Psi}\Psi$ it will include a piece given by $(\Psi^{0*}_- \Psi_+ + \Psi^*_+ \Psi^0_- )$.  The only operator in $\Psi_-^0$ which can operate in the physical subspace is the spurion, $\sigma_-$, so in the physical subspace the contribution to $P^-$ from this piece of $\bar{\Psi}\Psi$ is given by
$$
       \delta P^- \supset \mu \sigma_-^*\sigma_+ Z_-Z_+\int \normord{
e^{-i\sqrt{\pi}\tilde{\Sigma}(0,x^-)}} dx^- + C.C.
$$
Note that the product of spurions, $\sigma_-^*\sigma_+$, just acts like a c-number in the physical subspace.  Let us now apply mass perturbation theory to our construction.  We know that the wave function renormalization constants when expanded in a power series in $\mu$ have the forms
$$
Z_-(\mu) =  \sqrt{\fr{\chi e^\gamma},{2\pi}} \;+\; {\mathcal {O}}(\mu)\quad\;\quad Z_+(\mu) = \sqrt{\fr{m^2e^\gamma},{8\pi\chi}} \;+\; {\mathcal {O}}(\mu)
$$
If we take the unperturbed state to be a state of one Schwinger particle
$$
        \ket{p} \equiv \tilde{\Sigma}^*(p)\ket{\Omega(\theta)}
$$
we find that
$$
         \bra{p} P^+ \delta P^- \ket{p} = - 4 \pi\mu \bra{\Omega}\bar{\Psi}\Psi\ket{\Omega} = 2 m \mu e^\gamma cos\theta
$$
in agreement with many past calculations.

We must now notice a complication induced by the fact that the wave function renormalization constants must be calculated using Fermi products at space-like splitting.  $Z_+(\mu)$ and $Z_-(\mu)$ are determined by the relations
$$
           \{\Psi_+(x),\Psi_+(x + \epsilon_{SPACE})\} = \delta(\epsilon_{SPACE})
$$
and
$$
 \{\Psi_-(x),\Psi_-(x + \epsilon_{SPACE})\} = \delta(\epsilon_{SPACE})
$$
To evaluate these expressions we have to use $P^-$ to translate $\Psi_+$ off the surface $x^+ = 0$.  But $P^+$ depends on the wave function renormalization constants so the relations become implicit and a nonperturbative calculation would have to iterate, guessing a value for the renormalization constants then using the implied $P^-$ to check the values of the renormalization constants.

I believe that this somewhat stark conclusion is softened by the following observation.  The new piece of $P^-$ has the form
$$
\delta P^- \supset \# \int \Psi_+(bleached;regulated) dx^-
$$
where (in the physical subspace) \# is a c-number which we know in terms of the renormalization constants.  We know $\Psi_+(bleached;regulated)$; no iteration is required for it.  All the worst complications are necessary to determine \#.  So if we are willing to fit \# to some fact to which it is sensitive (perhaps a piece of data from two-dimensional experimentalists or possibly a symmetry as suggested by Simon Dalley~\cite{dalley}) we can avoid the need to iterate.  Other terms in $P^-$ involve the renormalization constants and similar considerations obviously apply to them.

Let us list here the mechanisms by which the features we have been studying come about in two dimensions:
\begin{itemize}
\item Zero-mode fields, one of which is a ghost, are needed to regulate $P^+ = 0$.

\item Completion of the space includes translationally invariant states which can mix with the vacuum.

\item Gauge invariance forces mixing to form a $\theta$-state, an eigenstate of the chargeless combinations of spurions.

\item The solution of the $\Psi_-$ constraint equation includes a solution to the homogeneous equation determined by $\{\Psi _{-}(x),{\Psi }_{-}^{\ast }(x + \epsilon_{SPACE})\} = {\delta }(\epsilon_{SPACE})$.

\item The solution to the homogeneous equation gives rise to an operator, proportional to the chiral condensate, which acts in the physical subspace.

\item Once that operator is in place, the problem can be solved in the usual light-cone space of   $\Psi_+$ (the unphysical, zero-mode fields are not further needed).
\end{itemize}

I shall make two further remarks on the two-dimensional case: Once the problem is formulated and the ``new'' term is included, one can discretize and use DLCQ to solve it; indeed, starting with DLCQ (but being careful to include $\Psi_-$) one can find the form of the ``new'' term but cannot correctly calculate the renormalization constants and so, cannot calculate the coefficient(\#)~\cite{mccartor99}.  If one fit the \#, one could solve the problem correctly and might never know that \# is proportional to a chiral condensate.  Finally, since the Schwinger particle has a nonzero mass for zero bare electron mass, the Schwinger model is a better model for the $\eta^\prime$ than the pion.  A better model for the pion is furnished by adjoint $QCD_2$ where we find $M(\mu)^2 \sim - \mu\langle\Omega |\bar{\Psi}\Psi|\Omega \rangle$.

\section{Four Dimensions}
We know that many of the ingredients which lead to the ``new''term in $P^-$ are also present in the case of four dimensions:
\begin{itemize}
\item In light-cone gauge a Lagrange multiplier field, $\lambda(x^+,x^\perp)$, must be present~\cite{bns}; that field is a zero norm field so the representation space is of indefinite metric.

\item The constraint equation which relates $\Psi_-$ to $\Psi_+$ admits an arbitrary function, $\Psi_-^0(x^+,x^\perp)$.  That function is determined by the requirement that the full $\Psi_-$ satisfy $\{\Psi _{-}(x),{\Psi }_{-}^{\ast }(x + \epsilon_{SPACE})\} = {\delta }(\epsilon_{SPACE})$ for {\it ANY} $\epsilon_{SPACE}$; the point of ANY is that there are space-like separations in the initial value surface (the $\perp$ directions) and this relation will normally be set by the initialization of $\Psi_-$; but it must also be satisfied for all space-like $\epsilon$'s which project onto z ($x^3$); thus the issue is really the isotropy of the $\Psi_-$ commutator.

\item The zero-mode fields will appear in exponentials in the Fermi fields (that was an important piece of the two-dimensional structure).
\end{itemize}
I do not know for certain if the completion of the space will include translationally invariant states but it appears to me that it will and I shall assume it does.

Even with these known aspects of the problem in hand I cannot give a complete analysis as I could in two dimensions.  Here I must proceed in part by analysis and in part by analogy.  In what follows I shall use the notation of the discrete case but the continuum is the same with the sums replaced in an obvious way by integrals.  We start by writing down a standard initialization of $\Psi^n_+(0,x^-,x^\perp)$ (n is a color index) but with the terms grouped in such a way as to allow us to factor each oscillatory mode in $x^\perp$

\pagebreak

\begin{eqnarray}
    \Psi^n_+(0,&&x^-,x^\perp) = \nonumber \\
&&{1\over\sqrt{\Omega}}\sum_{s,k_\perp}e^{ik_\perp x^\perp}\sum_{k_-} 
   b^n(k_-,-k_\perp,s) e^{-ik_-x^-} +
   d^{n*}(k_-,k_\perp,-s) e^{ik_-x^-}  \nonumber    
\end{eqnarray}
Similarly for $\hat{\Psi}^{n0}_-(x^+,x^\perp)$ (I shall come to the reason for the hat momentarily)
\begin{eqnarray}
   \hat{\Psi}^{n0}_-(x^+,&&x^\perp) = \nonumber \\ 
&&{1\over\sqrt{\Omega}}\sum_{s,k_\perp}e^{ik_\perp x^\perp}\sum_{k_+} 
   \beta^n(k_+,-k_\perp,s) e^{-ik_+x^+} +
   \delta^{n*}(k_+,k_\perp,-s) e^{ik_+x^+} \nonumber
\end{eqnarray}
The second sum in each of these expressions is the initialization of a two-dimensional Fermi field and can be bosonized by standard methods. For $\Psi^n_+$ we get
$$
 \Psi_+(0,x^-,x^\perp) = {1\over\sqrt{\Omega}}\sum_{n,s,k_\perp}e^{ik_\perp x^\perp}\sum_{k_-}e^{\lambda_+^{(-)}(n,k_\perp,s)}
\sigma_+(n,k_\perp,s)
e^{\lambda_+^{(+)}(n,k_\perp,s)}
$$
where
$$
\lambda_+^{(-)}(n,k_\perp,s) = -i\sqrt{{2\pi\over \Omega}}\sum_{k_-}
{1\over \sqrt{p_-}} C(k_-,n,k_\perp,s)e^{-ik_-x^-}    
$$
with
\begin{eqnarray}
&&C(k_-,n,k_\perp,s)  = \sum^{k_-}_{\ell} d^{n}(k_-,k_\perp,-s)\left(\ell\right) b^{n}(k_-,k_\perp,s)\left(k_\perp
- \ell + 1\right) + \nonumber \\ && \sum^{\infty}_{\ell} b^{n*}(\ell,k_\perp,s) b^n(k_- + \ell,-k_\perp,s)
- d^{n*}(\ell,k_\perp,-s) d^{n}(k_- + \ell,k_\perp,-s)\nonumber
\end{eqnarray}
We rewrite the expression for $\hat{\Psi}^{n0}_-$ in a similar way. For the case of the continuum, the fields $\lambda_\pm$ are Klaiber~\cite{kla} regulated.

Except for zero bare quark mass, $\hat{\Psi}^{n0}_-$ cannot be ${\Psi}^{n0}_-$.  That is because
$$
\{\hat{\Psi^{n0}} _{-}(x),{\hat{\Psi}^{n0\ast} }_{-}(y)\} = {\delta }(x^{+}-y^{+}){\delta }^2(x^{\perp}-y^{\perp})
$$
But, for nonzero bare quark mass $\Psi_-^{n}$ includes a functional of $\Psi_+^{n}$; indeed that is the only piece which is usually included.  Here I shall assume that for nonzero bare quark mass, the total $\Psi_-$ field remains canonical through a mixing of $\lambda$ with $\lambda_-$ in the exponential factors of $\Psi_-^{n0}$ but that the spurion, $\sigma_-(n,k_\perp,s)$, remains unchanged; that is what happened in two dimensions and it seems to be the only thing allowed by kinematics and the requirement that the equations of motion are satisfied in the physical subspace.  

We shall now assume that gauge invariance requires the physical vacuum to be an eigenstate of the products of spurions which carry the quantum numbers of the vacuum.  We have concentrated the (color) charge, the spin and the transverse momentum of the $\Psi_+^n$ field in the spurions.  We can therefore ``bleach'' the field by multiplying by the spurion from $\Psi_-^{n0}$ which carries the opposite quantum numbers.  Under the assumptions we have made there will be a term in $P^-$ which, in the physical subspace, acts as
$$
 \delta P^- \supset \# \sum_{n,s,k_\perp}\int e^{\lambda_+^{(-)}(n,k_\perp,s)}
e^{\lambda_+^{(+)}(n,k_\perp,s)} dx^-
$$
Here, $\# \sim \mu_q \langle\Omega |\bar{\Psi}\Psi|\Omega \rangle$, although if we choose to fit the \# as suggested above, we would not necessarily need to know that.  If the pion has zero mass for zero bare quark mass, this operator will give
$$
M_\pi^2 \sim \mu_q \langle\Omega |\bar{\Psi}\Psi|\Omega \rangle +\dots
$$

\section*{Acknowledgments}
This work was supported by the U.S. Department of Energy.

\end{document}